\documentclass[twocolumn]{aastex6}

\usepackage{graphicx}
\usepackage{natbib}

\def \be {\begin{equation}}
\def \en {\end{equation}}

\def \mta {}

\def \nm{}
\def \nn{}
\def \mm{}
\def \bbb{}
\def \zz{}

\def \vv{}

\def \vrr{}

\def \c {3C~279 }

\begin{document}

\title{
Meeting the Challenge from Bright and Fast Gamma-Ray Flares of 3C 279
}
 \author{
V.~Vittorini\altaffilmark{1}, M.~Tavani\altaffilmark{1,2,3,4},
A.~Cavaliere\altaffilmark{1,4} }

\altaffiltext{1}{INAF/IAPS-Roma, via Fosso del Cavaliere  100,
I-00133 Roma, Italy} \altaffiltext{2}{Universit\'a ``Tor Vergata''
Dipartimento di Fisica,  via della Ricerca Scientifica 1, I-00133
Roma,  Italy} \altaffiltext{3}{Gran Sasso Science Institute, viale
Francesco Crispi 7, I-67100 L'Aquila, Italy}
\altaffiltext{4}{Astronomia, Accademia Nazionale dei Lincei, via
della Lungara 10, I-00165 Roma, Italy}


\vskip .1in


\begin{abstract}

Bright and fast gamma-ray flares with hard spectra have been
recently detected from the blazar  3C 279, with apparent GeV
luminosities up to $10^{49}$ erg s$^{-1}$. The source is observed
to flicker on timescales of minutes with no comparable optical-UV
counterparts. Such observations challenge current models of
high-energy emissions from 3C 279 and similar blazar sources that
are dominated by relativistic jets along our line of sight with
bulk Lorentz factors up to $ \Gamma \sim 20$ launched by
supermassive black holes. We compute and discuss a model based on
a clumpy jet comprising strings of compact plasmoids as indicated
by radio observations. We follow the path of the synchrotron
radiations emitted in the optical - UV bands by relativistic
electrons accelerated around the plasmoids to isotropic Lorentz
factors $\gamma \sim 10^3$. These primary emissions are partly
reflected back by a leading member in the string that acts as a
moving mirror for the approaching companions. Around the
plasmoids, shrinking \emph{gap} transient overdensities of seed
photons build up. These are upscattered into the GeV range by
inverse Compton interactions with the relativistic electrons
accelerated in situ. We show that such a combined process produces
bright gamma-ray flares with minor optical to X-ray enhancements.
Main features of our model include: bright gamma-ray flares with
risetimes as short as a few minutes, occurring at distances of
order $10^{18} $ cm from the central black hole; Compton dominance
at GeV energies by factors up to some $10^2$; little reabsorption
from local photon-photon interactions.

\end{abstract}

\keywords{gamma rays: general, sources -- FSRQ Objects:
individual: 3C 279. }

\section{Introduction}

The Flat Spectrum Radio Quasar (FSRQ)3C 279 ($ z = 0.536$) is a
blazar prominent in  gamma rays. It was repeatedly detected above
100 MeV by EGRET \citep{hartman1992, kniffen1993}, AGILE
\citep{giuliani2009}, 
and \textit{Fermi}-LAT (e.g.,
\citealt{hayashida2012,hayashida2015}),
and also detected above 100 GeV by MAGIC \citep{albert2008}.
In 2015 June, \c was caught in a remarkably
bright flaring state by \citealt{ackermann2016}, hereafter A16.
The {\vrr apparent} gamma-ray luminosity attained $10^{49}$ erg
s$^{-1}$, (under isotropy assumption), with flux variability
timescales resolved down to {\nm 2-3} minutes and doubling times
of {\vv $\sim 5$} minutes.

In fact, the overall gamma-ray activity observed in \c  by
\citealt{hayashida2015} and  by A16 in 2015 June  shows strongly
enhanced radiation above 100 MeV lasting several days. Several
features are to be noted:  a very high "Compton dominance" in the
spectral energy distribution, {\mta marked by} a ratio of
gamma-ray to optical emission rising by factors 3 - 5  to attain
values $q \sim 10^2$ in a few hours; very rapid flickering with
resolved timescales down to a few minutes;  a flat gamma-ray
spectrum occasionally extending out to about  50 GeV
\citep{paliya2015}, 
with no evidence of local reabsorption. Such unprecedented
features set a challenge hard to meet for most current radiative
models of this source, and possibly of other blazars as well, such
as PKS 1222+216 \citep{aleksic2011}, 
PKS 1510-089 and 3C 454.3 \citep{coppi2016}. 

\section{Modeling the Gamma-Ray Source
}

Blazars  are dominated by a relativistic jet with a Lorentz boost
$ \Gamma \sim 10\,-\,20$, launched by  a supermassive black hole
(SMBH) along our line of sight. They  feature widely extended
non-thermal spectra,  which are interpreted in terms of
Synchrotron (S) emission and of Inverse Compton (IC) radiation
(cf. \citealt{rybicki1979});
these are produced by electrons
accelerated in the jet to attain random Lorentz factors up to
$\gamma_b \sim 10^3$. {\nn In particular, FSRQs} are often marked
by gamma-ray Compton dominance, though generally at less extreme
degrees than 3C 279 itself.

The S emission is observed at energies $\epsilon_s\propto
\gamma_b^2 \, \Gamma \, B'$ with a luminosity  $L_S \sim
c\,\sigma_T\, n' \, \ell'^3 \, \gamma^2_b\, U_B'\, \Gamma^2$.
These are given  in terms of the number density $n' \sim 10^3 $
cm$^{-3}$ of energetic electrons with $\gamma_b\sim10^3$, within
the source size  $\ell' \sim 10^{16}$ cm threaded by a magnetic
field of strength $B' \sim 0.1 - 1$ G with magnetic energy density
$U_B'=B'^2\,/\,8\pi \simeq 10^{-2} \, \rm erg \, cm^{-3}$ (primed
quantities refer to the comoving frame). Such emission is widely
held to explain the continuum observed from the IR to \emph{UV
}bands. IC scattering is often taken into account for the
gamma-ray yield {\mta from the same} electron population. This
operates on any density $U'$ of {\nm soft}``seed" photons present
in the jet by conserving their number while
upgrading their energy 
to yield observed luminosities {\vv $L_{IC} \sim c\,\sigma_T\,n'\,
\ell'^3 \, \gamma^2_b \,U' \, \Gamma^2$}.

On the other hand, all FSRQs also share with the other quasars two
thermal features: strong, nearly isotropic broad emission lines
shining in the optical band and produced in the Broad Line Region
(BLR); a bright Big Blue Bump {\nn (BBB)} comprising convolved
continua produced by the hot inner rings of the accretion disk
surrounding the central SMBH (e.g.,
\citealt{peterson1997, peterson2006}). 
To wit, 3C 279 features all radiative components observed in FSRQ
blazars, but during its flares the gamma-ray band is enhanced and
variable to an extreme degree.

In the  present  Letter we discuss the implications of these
findings, and
propose an interpretation of the gamma-ray flaring activity of \c
in terms of a clumpy jet. The latter  features strings of
plasmoids as long recognized in the radio observations of this and
other blazars (see \citealt{hovatta2009} and references therein).
Indeed, radio observations of plasmoid kinematics in \c
\citealt{hovatta2009}
indicate  a 
boost $\Gamma \simeq 20$ that we adopt in the present paper. {\vrr
In our model the jet has an opening angle $\sim\, \Gamma^{-1}$}.

The short time scales $t_v \sim $ a few minutes observed in the
{\mta June 2015} gamma-ray flares of \c {\vv point to emission
from regions of small proper sizes $d' \lesssim c\,t_{v} \;
\Gamma^2 \simeq \rm 10^{13} \, cm \,\, \Gamma^2 $}. High
luminosity coupled with fast variability constitutes the challenge
to be met by a viable source model.

A guide toward a satisfactory IC  model  is provided by
considering the energy density of the  seed photons after the
pattern \be U' = \frac{L'}{4 \pi c d'^2}  \; , \label{eq-guide}
\en in terms of the \emph{effective} luminosity $ L' $ of their
source and of the size $ d'$  of the containing volume. It is seen
that smaller size values not only enhance $U'$, but also shorten
the variability time scale, so as to achieve a substantial Compton
dominance  {\nm within} short times. Meanwhile, smaller sizes
decrease the local reabsorption of the IC radiation by
photon-photon {\nn pair-producing} interactions. In fact, our
model focuses on physical conditions where small values of $d'$
arise.

\section{IC Radiation from  BLR Seed Photons? }

Meeting the above requirements  while retaining  the bound $\Gamma
\lesssim 20$ for \c  make  the models limited to seed photons from
the BLR problematic.  Such  models are based on combinations of S
and IC radiated at radii $R \lesssim 0.1 $ pc by a population of
relativistic electrons with Lorentz factors $\gamma \lesssim 10^3$
as discussed by \citealt{hayashida2015} and A16.
The electrons
inhabit the jet and interact with the local magnetic field $ B'
\lesssim 1$ G to produce S emission, {\nn while} by the IC process
upscatter \emph{any seed} photons of density $U'_{BLR}$ as seen in
the jet frame.

In the Synchrotron-Self Compton approach (SSC, e.g.,
\citealt{maraschi1992}), 
the very \emph{same} S photons are upscattered into the GeV range,
to yield a basal level of IC $\sim $ S {\nn luminosities} {\mta
resulting in $q\sim 1$}. External radiation Compton (ERC) {\mta
may} yield larger IC {\mta emission } by upscattering
\emph{additional} UV photons produced \textit{in the BLR} at
$R_{BLR} \sim 0.1$ pc by BBB light reprocessed/reflected by {\mta
gas clouds} that float around at some $10^3 \, \rm km \, s^{-1}$
within the region. The {\mta resulting} energy density is given by
\be U'_{BLR} \simeq (7\times 10^{-3} \; {\rm erg \; cm^{-3}})\,
\frac{\xi}{0.02} \, (\frac{R_{BLR}}{0.1\,{\rm pc}})^{-2} \, L_{D,46}  \,
\Gamma^2 \,  , \label{eq-1}\en where we assumed a disk luminosity
$L_D = 10^{46} \, \rm erg \, s^{-1}$ appropriate for 3C 279, a BLR
covering factor $a = 0.2$ and a  cloud reflectivity $f_c  \simeq
0.1$ to produce $\xi = a \, f_c \simeq 0.02$. Such a value for
$U'_{BLR}$ may yield a ratio {\nm IC to S luminosities} $q =
U'_{BLR}/U'_B \simeq 50$, still insufficient for explaining the
2013 December  {\mta and 2015 June} bright and fast gamma-ray
flares of 3C 279.

On route toward higher  IC/S ratios,  one may consider the more
abundant IR photons that are radiated by the dusty torus around
the accretion disk upon reprocessing the BBB light; however, such
photons  are less effective than the UV ones for upscattering into
the GeV range (see \citealt{boettcher2013}). 
Alternative scenarios just make do with lower values of
$B$ and modest accelerations,
but they have to assume  $\Gamma \,>\, 35 $  
to avoid reabsorption of the meager IC radiation by photon-photon
interactions within {the BLR } (\citealt{hayashida2015}, A16).

Straightforward consequences of these models include a very
\emph{low} magnetic field $B\,<\,0.1\,$G {\nn in the jet} and the
{related} bulk magnetization $\sigma_j \propto B^2/\Gamma^2 \sim
10^{-4}$ (A16). This {\mta would imply} (see e.g.,
\citealt{mignone2013, yuan2016, cavaliere2017}) 
it has a jet that resists to forming  MHD structures that are
conducive to driving magnetic reconnections and
micro-instabilities active for electron accelerations.  In
addition, these - if they somehow were driven - would be
ineffective to attain high values of $\gamma \simeq 10^3 $ (see
next Sect.4). Alternative scenarios {\nn for FSRQ emissions}
consider special structures, such as spine-sheaths
\citep{tavecchio2008, sikora2016} and "rings of fire"
\citep{macdonald2015}.

We conclude that achieving substantial  Compton dominance $ q \sim
100$ coupled with very fast variability requires  the source to
have substantially \emph{higher} values of $U'$ \textbf{compared
with} Eq. \ref{eq-guide} within far smaller volumes than can be
provided
by the BLR, as anticipated in Sect. 2 and 
initially discussed in \citealt{vittorini2014}. 
{\vrr Moreover, the hard unabsorbed gamma ray spectra up to $10^2$
GeV often observed in such sources (see, e.g.,
\citealt{costamante2017}) point toward emission sites beyond the
BLR.}

\section{Enhanced  Seed Photons from  Moving Mirrors}

\begin{figure*}
\begin{center}
\includegraphics[width=13cm]{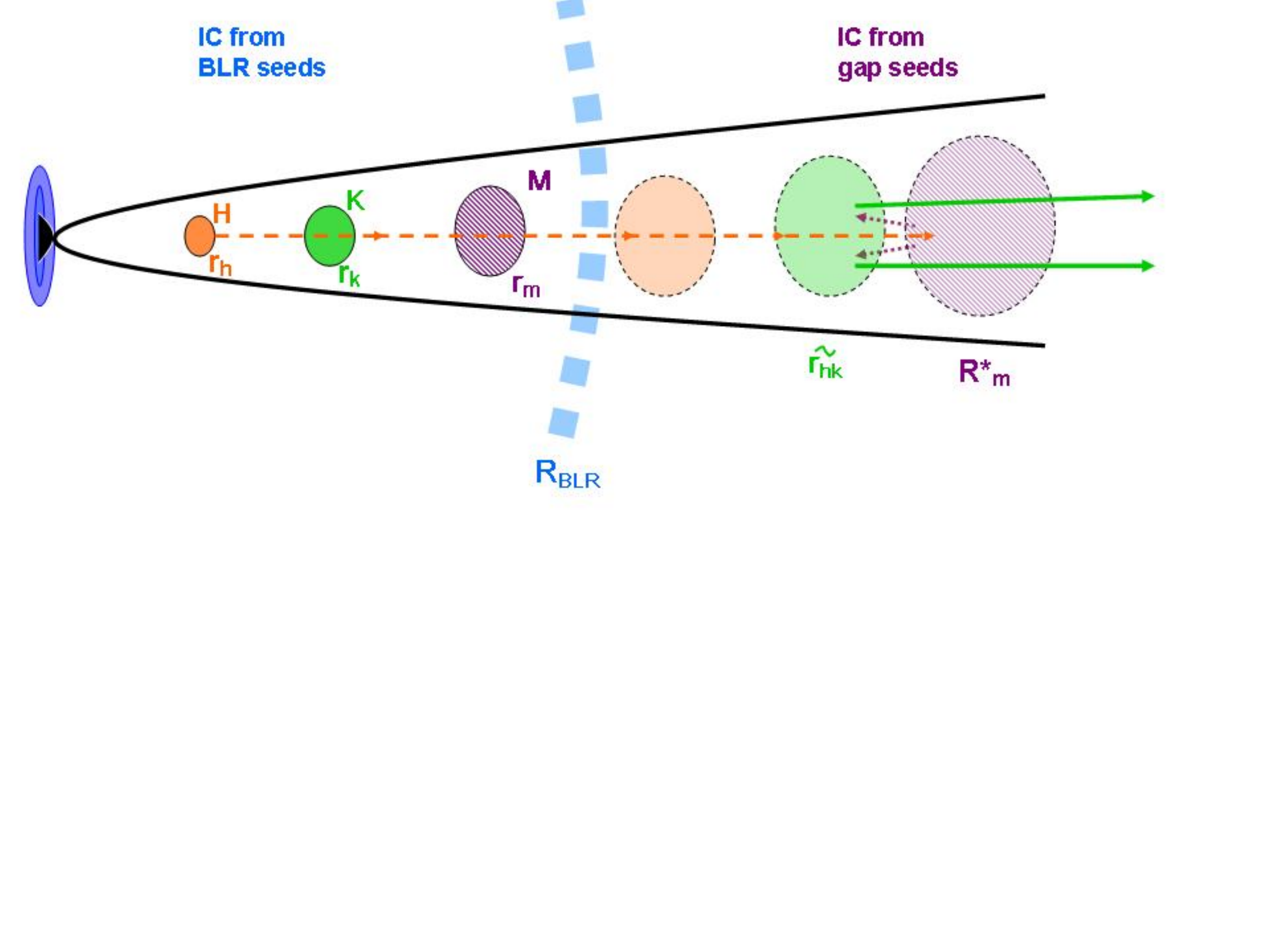} 
\vspace*{-5.cm} \caption{ {\mm Schematic picture } 
of the source structures we expect at and beyond the BLR. S
radiation (orange dashed line) is emitted by inner plasmoids such
as $H$ at a radius $r_h$, and is reflected back (violet dotted
lines) by the outermost mirror-plasmoid $M$ at a radius $R*_m$
toward an incoming plasmoid {\vv $K$: the encounter occurs at a
radius $\tilde{r}_{hk}$}. Dense seed photons build up in the gap
between the approaching plasmoid $K$ and the mirror, where they
are IC upscattered to gamma-ray energies (green dotted lines) by
electrons accelerated in and around plasmoids.}
 \label{fig-1}
\end{center}
\end{figure*}

Thus, we are led to develop the scenario proposed in
\citealt{tavani2015}.
In the present Letter, we base our
model on a clumpy jet with a bulk magnetization
$\sigma_j\gtrsim1$; the jet comprises a sequence of plasmoids
moving out  with speeds $\beta$ and boosts $\Gamma \leq 20$. These
plasmoids are related  to tearing instability and reconnections of
\textbf{B}-lines in the jet that are squeezed or even invert their
polarity within a collisionless plasma layer. Such processes have
been widely proposed, discussed, and numerically computed (cf.
\citealt{kagan2015, coppi2016, yuan2016,
lyutikov2016,petropoulou2016}; see also \citealt{burch2016} for
{\vv observational} evidence in an astrophysical context).
They form \emph{strings} of separate "magnetic islands" that
proceed to merge into  larger and larger plasmoids. We refer to
plasmoid sizes $\ell' \sim 10^{16}$ cm at a distance $R < 0.1$ pc
from the central SMBH. Their origin is related to the changing
topology of the \textbf{B}-lines that induces \textbf{E} fields
localized
between plasmoids 
{\nn in} gaps {\nm of around} $10^{16}$ cm, and shrinking. These
\textbf{E }fields are known from a number of detailed kinetic
simulations to \emph{accelerate} electrons up to $\gamma \sim
10^3$ in conditions where the {\mta local} electron magnetization
is \emph{large}, i.e.,  $\sigma_e > \sigma_j \, m_p/m_e \sim
10^2$.

The relative plasmoid distances in a string vary depending on
their different speeds along the jet, with the leading one
significantly slowed down
primarily  by snow-plow effects 
against jet material, in particular, BLR clouds. This constitutes
an  interesting condition as the leading member in a string can
reprocess and partially reflect {\mta some of the primary} S
radiation emitted {\nn by the trailing members} (photons
''mirrored" by slow clouds in the BLR were originally proposed  by
\citealt{ghisellini1996}).  

Shrinking mirror-plasmoid \emph{gaps} so produced provide {\it
compact} sites with {\it high} seed photon densities for intense
IC to take place. In a scenario with  multiple emitting plasmoids
(see Fig. 1), the primary S flux in {\nn the optical-UV} bands is
emitted by a plasmoid like $H$ starting within the BLR, that is,
in the {\nn range} $0 < r_h <R_{BLR}$) The S flux is emitted {\nm
forward} and mirrored backward by the leading 
{\nm plasmoid} $M$ {\nn toward} an incoming plasmoid $K$. The
whole process is subject to a ''causality condition"
(\citealt{boettcher1998}); 
the mirrored photons are received by {\nm an} incoming plasmoid
when the latter is at the receiving point {\nm that} we denote
with $\tilde{r}_{hk}$ ({\nn cf.} Fig. 1).

The  condition is easily evaluated when the same plasmoid moving
with radial coordinate $r=r_0+\beta ct$ and boost $\Gamma$ first
emits 
{\nm at} $r$, and then receives the radiation reflected backward
by the mirror moving with coordinate $r_m=r_{mo}+\beta_m ct$ and
boost $\Gamma_m$.  The receiving point reads
$\tilde{r}=\tilde{r}_0+r[(1+\beta_r)\Gamma_r]^{-2}$ in the
AGN-frame, for primary photons emitted in the named range
{\vv and reflected back  at
$R^*_m=R^*_{m0}+\beta_m(1-\beta)r\,/\,[\beta(1-\beta_m)]$}. Here
$\beta_r=(\beta-\beta_m)(1-\beta \beta_m)^{-1}$ and
correspondingly $\Gamma_r=\Gamma \Gamma_m(1-\beta \beta_m)$  are
the relative velocity and boost between the receiver and the
mirror (note that
$\Gamma_r\simeq\Gamma\,/\,[\Gamma_m(1+\beta_m)]$).  {\nm Moreover,
$\tilde{r}_0=2(\beta r_{m0}-\beta_m
r_0)[(1+\beta)(1-\beta_m)]^{-1}$ is the receiving point} for
primary photons emitted at $r\simeq 0$ and reflected at
$R^*_{m0}=(\beta r_{m0}-\beta_m r_0)/[(1-\beta_m)\beta]$. Note
that for small values of $r_0$ the approximation $R*_{m0}\simeq
2\Gamma_m^2\,r_{m0}$ holds; so the factor $\Gamma_m^2$ makes it
possible to have relevant reflections \emph{beyond} $R_{BLR}$.
The IC {\nm radiative} event is completed when the plasmoid $K$
crosses the {\bbb effective} gap defined by
 \be d_{g}\equiv \tilde{r}\,-\,\tilde{r}_0\,=\,\frac{R_{BLR}}{(1+\beta_r)^2 \, \Gamma_r^2}
\, ,\label{eq-gap} \en and lasts for an observer time $t_v =
d_{g}/[c\beta(1+\beta)\Gamma^2]\simeq R_{BLR}/[8 \, c \, \Gamma^2
\, \Gamma_r^2]$: this {\nm turns} out to be about {\nm 2 minutes }
when, e.g., $R_{BLR}\simeq 3 \cdot 10^{17} \, \rm cm$, $\Gamma
\simeq 20$, and $\Gamma_r \simeq 5$ hold.

In the initial stage, the primary emitter radiates from an inner
position $r \ll R^*_m$ and the mirror surface $\pi\ell'^2_m$ is
small relative to the
{\bbb cross section of the emission beam}
 $\pi R^{*} _{m}\, ^{2}\, \Gamma^{-2}$
as to reduce the \emph{effective} power received by the mirror {to
 $4 \, L'_S\,[\ell'_m\,/ \,(R^*_m-r)]^2\,\Gamma_r^4 $ in the    
head-on approximation and  with $ 1+\beta_r \simeq 2$}. A fraction
{\vv $f_m\simeq n'_{cold}\sigma_T\ell'_m$} thereof is reflected
back and is received by the incoming plasmoid when the latter is
close to the reflection point, that is, in a {cone  of cross}
section of order $\pi\ell'^2_m$. Then, in the gap we have
\begin{equation}\label{far}
U'_m \simeq  (5\,10^{-5} \, {\rm  erg \, cm^{-3}})\frac{\,f_{m,-1}
\, L'_{S,43}\, \Gamma_r^{6}}{(R^*_m-r)_{18}^{2} } \, .
\end{equation}
The denominator represents the  beam dilution of the primary
photons, while the volume {\zz swept} per unit time is
$\pi\,c\,\ell'^2_m$. Such values of $U'_m$ are still insufficient
to dominate over $U'_{BLR}$ as given by Eq. \ref{eq-1}; in
addition, for \c flares very short risetimes are required by A16
observations.

As  the emitting plasmoid travels toward and beyond $R_{BLR}$, the
mirror surface  fills up the primary emission cone, and the
\textit{effective} received power saturates to {\nn the full
value} $L'_S \, \Gamma_r ^2$.  The fraction {\vv $f_m$}  is
reflected toward the incoming plasmoid {\nn until} the receiving
radius $\tilde{r}$ becomes very close to the reflection point.
Thus, high-energy densities of seeds {\mm prevail} in the comoving
frame,  and   for primary photons emitted at $r\simeq R_{BLR}$
attain a maximum that reads \be U'_m \lesssim  (2 \cdot 10^{-1} \,
{\rm \, erg \, cm^{-3}})\,\frac{f_{m,-1} \, L'_{S,43}\,
 \Gamma_r^4}{\ell'^{2}_{m,16}} \, \label{eq-3} . \en

We may compare  the above Eqs. \ref{far} and \ref{eq-3} with the
photon energy density experienced by a plasmoid crossing the BLR,
that is, with $U'_{BLR} $ given by Eq. \ref{eq-1}.
Remarkably, if $ \Gamma \simeq 20$ holds, 
we have
 $U'_m > U'_{BLR} $ for $ \Gamma_r \simeq 5$ (corresponding to
 $\Gamma_m \simeq 2$); 
with 
{\nn such} values an interesting configuration  is obtained
 {\mm in the {moving} mirror scenario}.
We stress that high seed photon {\nn densities} {confined} to the
gap
 may well occur beyond the BLR,
{\mta a welcome feature of our model that minimizes reabsorption
by local photon-photon interactions.}

\section{Simulations}

Our  simulations treat in  detail  plasmoids ejected in {\mm a
string} along
the jet with different velocities. Each 
{\mm member} $H$ in the string has proper size $\ell' _h$ and
moves along the jet according to $r_h=r_{h0}+c\beta_h t$, $r_h$
being the distance from the SMBH at time $t$ in the AGN frame.
{\mta Key elements are: (1) non-uniformity of the plasmoid boosts in the
string,
with {a slower}  leading member;
and (2) 
 {\mm the leader acts} as a moving mirror that partially
reflects back S radiation toward an approaching companion $K$.}

We numerically simulate the emissions resulting from a set of
values $\Gamma_h$, for $h$ running from 1 to $6$, {\nn with the
upper value} denoting the mirror; we use the {\mm subscripts} $h$
for the primary emitter, $k$ for the receiver, and $m$ for the
mirror. {\mta For simplicity, we carry out the formalism in the
AGN reference frame where $R^*_{hm}$ is the distance from the
central BH when the photons emitted at $r_h$ by {the plasmoid $H$}
are reflected.}

The primary S emission $L'_{S,h}$ from $r_h$ occurs at the time
$t_h=(r_h-r_{h0})/(\beta_h \, c)$, and is reflected back by the
moving mirror at the time \be t_h^*=\frac{\beta_h
r_{m0}-r_{h0}+r_h(1-\beta_h)}{c\beta_h(1-\beta_m)}
\label{eq-t*}\en when the mirror is at the reflection point
$R^*_{hm}\equiv r_{m0}+c\beta_m t_h^*$. We {\mta then} take into
account the causality condition for the receiving plasmoid $K$;
accordingly, mirrored photons are received back at the time \be
\tilde{t}_{hk}=\frac{R^*_{hm}-r_{k0}+ct_h^*}{c(1+\beta_k)} \; ,
\label{eq-reenter}\en when the receiving plasmoid $K$ is at
$\tilde{r}_{hk} = r_{k0}+c\beta_k\tilde{t}_{hk}$. Here, the
mirrored photons are IC upscattered to gamma rays. The primary
emission occurs in the {\mm range}
$0<r_h<R_{BLR}$; for $r_h$ spanning this 
{\mm range} we obtain the related receiving points, subject to
the obvious condition $\tilde{r}_{hk} < r^*_k$, in terms of
the {\nm radius} where plasmoid $K$ touches the mirror.

At the receiving point, the electron population of {plasmoid}
$K$ upscatters after the IC processes the mirrored photons {\mta originally emitted by} 
{\nm plasmoid} $H$. The energy density of these seed photons in
the {\mta IC} scattering region is
\begin{equation}
U'_{hk}\simeq\frac{f(1-f)^{2N-h-k}L'_{s,(h)}\,
[(1+\beta_{r,h})\Gamma_{r,h}]^4\,\Gamma_{r,k}^2}{4\pi
c\,(R^*_{hm}-r_h)^2
} \, ,\label{eq-U}
\end{equation}
where $N$ is the number of emitters in our
string,  $\Gamma_{r,h}$ are the relative boosts 
{\mm of the  plasmoids relative to the mirror}; note that
$U'_{hk}$ can grow until $r_h$
{\nm becomes comparable with} $R_{BLR}$, as anticipated in Sect. 4
and detailed by Eq. \ref{eq-3}.

In Fig. 2 we present {\mta an example (with $N=5$ plus the mirror)
of our simulated gamma-ray light curves} that result from the sum
of all mirroring events and {\mm include} the contribution of the
standard IC from the
BLR. Note that 
{\nm such light curves} feature several very \emph{bright} and
\emph{short} spikes rising on timescales of a few minutes, as
fit to {\mta explain the A16 observations of 3C 279.}

\begin{figure}
\vspace*{-1.cm}  \hspace*{0.1cm}
 \includegraphics[width=19cm]{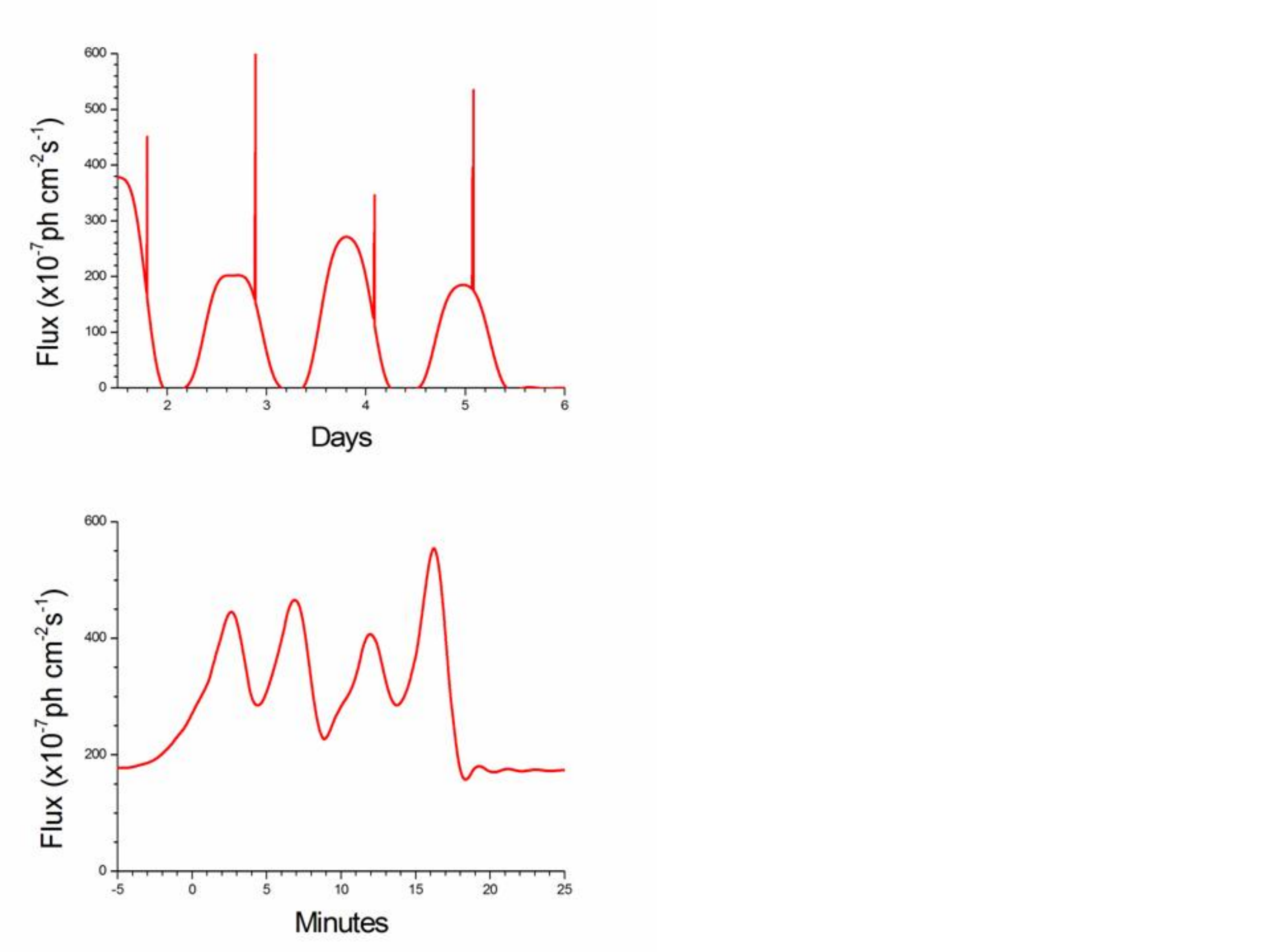}   
\vspace*{-.1cm} \caption{Gamma-ray light curves from our
simulations based on the moving mirror model for the gamma-ray
source in 3C 279. {\it Upper panel:} the simulated light curve
over a several-day timescale to be compared with Fig. 1 of A16.
The emission spikes are the result of
minute-timescale variability that is smoothed out for integration
time bins of a few hours as used by A16.
 {\it Lower panel:} simulated light curve
calculated for short timescales of the order of minutes, to be
compared with Fig. 2 of A16. 
}
 \label{fig-1}
\end{figure}

\section{Discussion and Conclusions}

The super-bright, super-fast, {hard spectrum} gamma-ray flares
from 3C 279 {\nm in 2015 June} call for a strong, compact source
beyond the BLR. These requirements challenge radiative processes
and source structures proposed over the years in the context of
blazar emissions. Features of our model that meet the challenge
include:
\begin{itemize}
\item {\it Power and Compton dominance}. Very intense and fast
gamma-ray flares are produced by the IC process {\mta {\it
beyond}} the BLR, arising from  strongly enhanced density of
reflected  {\nm seed} photons {\mm that are} localized in
mirror-plasmoid \emph{gaps} along the jet. In fact, {\mm based on}
the photon energy density $U'_m$ of Eq. \ref{eq-U}, we {\mm
derive} the energetics of the Compton-dominant flare and its
requirements.
{\mm To account for} the 
{\nm intrinsic} gamma-ray power $L_{\gamma}\sim 2 \cdot
10^{46}$erg s$^{-1}$, we {\nm need} an IC luminosity $L'_{IC}
\simeq 5 \cdot 10^{43} \, \rm erg \, s^{-1}$ in the comoving
frame. Since $L'_{IC} \simeq n' \, (4\pi/3)\, \ell'^3 \, \sigma_T
\, c \, U' \, \gamma_b^2$ holds, we require number densities of
radiating electrons $n' \sim \, 10^{2}\,\div\,10^3 \, \rm cm^{-3}$
with Lorentz factors up to $\gamma_b \, \sim 10^3$, and seed
energy densities $U'_m \sim 10 \, \rm \, erg \, cm^{-3}$ within
the gaps {\nm that considerably exceed those prevailing in the BLR
after Eq. 2}.

\item {\it Correlations}. The {\mta moving} mirror mechanism
naturally {decouples} the IC gamma-ray radiation produced in the
gaps from the optical-UV emissions prevailing inside the BLR.
{\vrr This decoupling occurs on timescales $\simeq 2(R^*_m\,
-\,\tilde{r})\,/\,c\,\sim$ half a day}, so very large gamma-ray
flares with Compton dominance $q \sim$ some $10^2$ {\mta can be
produced} {\mm over and above a slowly varying background,} as
observed in 3C 279.

\item {\it Short timescales}. We follow the photon paths {\nm as
shown in Fig. 1}: from S emission toward the mirror and backward
from it. We obtain the  risetime
 \begin{equation}
 t_v = (1+z) \,  R_{BLR} \, / (8c \,\Gamma^2 \Gamma_r^2).
 \end{equation}
In our conditions given by  $ \Gamma\simeq 20$, $\Gamma_m \simeq
2$ and $R_{BLR} \simeq 3\cdot 10^{17}$cm,  
{\mta this  comes to } about  {\nm 3 minutes}, including the
slowing  down by the source redshift effect {\mm at $z = 0.536$}.

\item {\it Hard, unabsorbed  gamma-ray spectra}.
We recall from the beginning of Sect. 4 that in the present
context electron accelerations take place in the very same narrow
\textit{gaps  $d_g$} (cf. Eq. \ref{eq-gap}) where the highest seed
density is built up and the induced $\mathbf{E}$-fields are
largest. So the electrons can replenish their fast IC losses and
retain a flat energy distribution. {\vrr On the other hand, the IC
upscattering of the mirrored photons occurs at radii beyond
$R_{BLR}$;} there we evaluate the optical depth to photon-photon
pair-producing interactions in the AGN frame from
$\tau_{abs}\simeq \sigma_T \, U_m \, d_g/(3\,\epsilon)\, \sim 0.2
\,$ (where $\epsilon\sim 1$ eV is the typical energy of the {\nm
soft} target photons) and obtain values smaller than unity
even with conservative values $\Gamma \sim  20$.
\end{itemize}

{\vrr We assume the plasmoids move along the jet following its
opening angle $\theta\,\sim\, \Gamma^{-1}$, so the transversal
size is $l'\,\simeq\,\theta\,R\, \simeq 10^{16}$cm at the BLR edge
and becomes a few times larger at the receiving point
$\tilde{r}\,\sim\,5\,10^{17}$cm. The \textbf{angular} alignment
between  \textbf{plasmoids and mirror $ l'_m/\tilde{r} \sim 3 \,
10^{-2}$} implied by the model  yields rare strong flickers over
and above a much longer plateau, as indicated by the
observations.}

We {\nm note} that the energetics required by the 2015 June
\textbf{event} observed 
from \c that attained $E\simeq L'_{\gamma}\,\Gamma^2\,t_v \,\simeq
\, {\rm a \, few}  \, 10^{48}$ erg
can be matched by a {\mm relativistic} energy content of a
plasmoid given by $E\simeq (4/3)\, \pi \,\ell'^3 \, n' \, m_e \,
c^2 \, \gamma_b \, \sim$ a few $10^{48}$ erg, with $n' \sim 10^3
\, \rm cm^{-3}$.  This implies a pressure balancing that of the
magnetic field and can be maintained by \textit{continued}
accelerations from magnetic reconnections, as computed and
discussed by \citealt{petropoulou2016}. 

On the other hand, a mirror Compton reflectivity given by $f
\simeq n' \, \sigma_T \, l'_m \, \sim $ a few \% requires electron
densities in the \textit{mirror} around a few $ 10^6$ cm$^{-3}$;
this leads to a {\vrr proton-dominated kinetic power $\sim\,\pi\,
l^2_mc\, n'm_pc^2\Gamma^2_m\,\sim\,10^{47}$erg s$^{-1}$ of the
mirror. Note that deceleration and snow-plough effects through the
jet will affect the leading mirror and help it to slow down toward
small values $\Gamma_m\,\sim$ 2 (see end of Sect. 4). Such low
values ensure a mirror contribution to the jet power of the same
order (or slightly below) that of the other plasmoids, in spite of
its higher density. So the \textit{total} power carried by the jet
is $L_j\,\sim\,10^{47}$erg s$^{-1}$ close to the Eddington limit
for a BH mass of $10^9M\odot$. Such a value is at the upper end of
the FSRQ range discussed  by \citealt{celotti2008}.} 

Virtues of {\nm our} model include: boost values  within
$\Gamma \simeq 20$, as provided by the radio observations {\zz of 3C 279}; standard 
{\nm values of the} magnetic field
 $B \sim 1 $ G;   electron magnetization
$\sigma_e > 
10^2$, 
high enough as to allow efficient electron
acceleration up to $\gamma \sim 10^3$ {\mm around 
or within plasmoids}; and apparent gamma-ray luminosities  up to
$10^{49}$ erg s$^{-1}$ from {compact } regions.\\
\\
We acknowledge partial support trough the ASI grant
no. I / 028 / 12 / 2.

\bibliography{biblio2}{}
\bibliographystyle{aa}

\end{document}